\documentclass[prl,twocolumn,showpacs,floatfix]{revtex4}
\usepackage{amsmath}
\usepackage{amssymb}
\usepackage{graphicx}
\usepackage[english]{babel}
\usepackage{latexsym}
\usepackage{graphics}
\usepackage{subfigure}

\def\be{\begin{equation}}
\def\ee{\end{equation}}
\def\bea{\begin{eqnarray}}
\def\eea{\end{eqnarray}}
\def\bse{\begin{subequations}}
\def\ese{\end{subequations}}

\def\be{\begin{eqnarray}}
\def\ee{\end{eqnarray}}

\begin{document}

\title{Mean-field Dynamics of Spin-Orbit Coupled Bose-Einstein Condensates}
\author{Yongping Zhang}
\author{Li Mao}
\author{Chuanwei Zhang}
\thanks{Corresponding author, email:cwzhang@wsu.edu}

\begin{abstract}
Spin-orbit coupling (SOC), the interaction between the spin and momentum of
a quantum particle, is crucial for many important condensed matter
phenomena. The recent experimental realization of SOC in neutral bosonic
cold atoms provides a new and ideal platform for investigating spin-orbit
coupled quantum many-body physics. In this Letter, we derive a generic
Gross-Pitaevskii equation as the starting point for the study of many-body
dynamics in spin-orbit coupled Bose-Einstein condensates. We show that
different laser setups for realizing the same SOC may lead to different mean
field dynamics. Various ground state phases (stripe, phase separation, etc.)
of the condensate are found in different parameter regions. A new
oscillation period induced by the SOC, similar to the Zitterbewegung
oscillation, is found in the center of mass motion of the condensate.
\end{abstract}

\affiliation{Department of Physics and Astronomy, Washington State University, Pullman,
WA, 99164 USA}
\maketitle

%03.75.Kk Dynamic properties of condensates, collective and hydrodynamic excitations, superfluid flow.
% 03.75.Mn. Multicomponent condensates; spinor condensates.
% 37.10.Vz Mechanical effects of light on atoms.
% 71.70 Ej Spin-orbit coupling, Zeeman and stark splitting, Jahn-Teller effect

Spin-orbit coupling (SOC) for electrons plays a crucial role in many
important condensed matter phenomena and applications, such as anomalous and
spin Hall effects \cite{AHE}, topological insulator \cite{TI}, spintronic
devices \cite{Zutic}, \textit{etc}. However, the observation of SOC physics
in nature solid state systems are often hindered by the unavoidable disorder
and impurities. In this context, ultra-cold atomic gases provide an ideal
platform for exploring novel SOC physics and device applications, owing to
their unprecedented level of control and precision in experiments. In
ultra-cold atomic gases, SOC can be generated through the laser-atom
interaction, which yields Abelian or non-Abelian gauge fields for atoms in
the dressed state basis \cite{Ruseckas,Zhu,Liu}. The recent broad interest
in spin-orbit coupled cold atoms is mainly motivated by their remarkable
applications \cite%
{Ruseckas,Zhu,Liu,Stanescu,Juzeliunas,Vaishnav,Stanescu2,Wu,Zhang1,Liu2,Spielman1,Lin,Zhang,Ho,Wang,Merkl,Larson}%
, ranging from the observation of Majorana fermions and the associated
non-Abelian quantum statistics \cite{Zhang1,Zhang}, the design of
atomtronics/spintronics devices \cite{,Juzeliunas,Vaishnav,Seaman}, to the
generation of magnetic monopoles \cite{Ruseckas}, \textit{etc}. An important
difference between electrons and cold atoms is that, while electrons are
fermions, ultra-cold atoms may be bosons, leading to novel spin-orbit
physics that has not been explored in solids.

A benchmark experiment along this research direction is the recent
realization of the one-dimensional (1D) SOC for cold bosonic atoms \cite{Lin}%
, which brings a completely new avenue for the study of the many-body
dynamics of spin-orbit-coupled Bose-Einstein condensates (SOC-BECs). It is
well-known that the starting point for the investigation of many-body
dynamics of a BEC is the mean-field Gross-Pitaevskii (G-P) equation \cite%
{Pethick}, whose general formula is still lacked for the SOC-BEC. In this
Letter, we derive a generic G-P equation for the SOC-BEC and investigate
their ground states and collective excitations. Our main results are:

(i) We find that the mean field dynamics depend on not only the SOC itself,
but also the method to generate it because the effective pseudospin states
in different laser setups for the realization of the same SOC can be
different superpositions of atom hyperfine states with different \textit{s}%
-wave scattering lengths. This fact is\ now taken into account in the
mean-field interaction terms in the pseudospin space in the G-P equation. We
find that the mean-field interaction energy may depend not only on the
density, but also the phase of the condensate (\textit{i.e.}, terms like $%
\Psi _{1}^{2}\Psi _{2}^{\ast 2}+c.c.$), which, to the best of our knowledge,
have not been explored in previous literature \cite{Wang,Merkl,Larson}. We
show that the phase dependent mean field terms can emerge for one type of
laser setup (denoted as a complex system), but vanish for another (denoted
as a simple system), with both laser setups implementing the same Rashba SOC.

(ii) We analyze the condensate wavefunctions for both systems in various
parameter regions. In the strong interaction and SOC region, there exist two
distinct phases for the condensate density: Thomas-Fermi (TF) and stripe.
While in the weak\ and medium interaction or SOC region, spatial separation
between two pseudospin components is observed.

(iii) The low energy collective excitations in spin-orbit coupled BECs are
investigated, for the first time, through the center-of-mass (COM) motion of
the condensate when a sudden shift of the center of the harmonic trap is
applied. We find a novel, shift direction and distance dependent,
oscillation frequency in the COM motion induced by the SOC, similar as the
Zitterbewegung (ZB) oscillation in the free space \cite{Vaishnav}. However,
the new oscillation period is linearly proportional to the\ SOC strength (in
contrast to the inverse proportion in the ZB oscillation \cite{Vaishnav}).
It is also independent on the atom interaction strength. Both oscillation
period and amplitude are different for the simple and complex systems, even
though they share the same physical parameters.

\begin{figure}[t]
\label{sch} \includegraphics[width=1\linewidth]{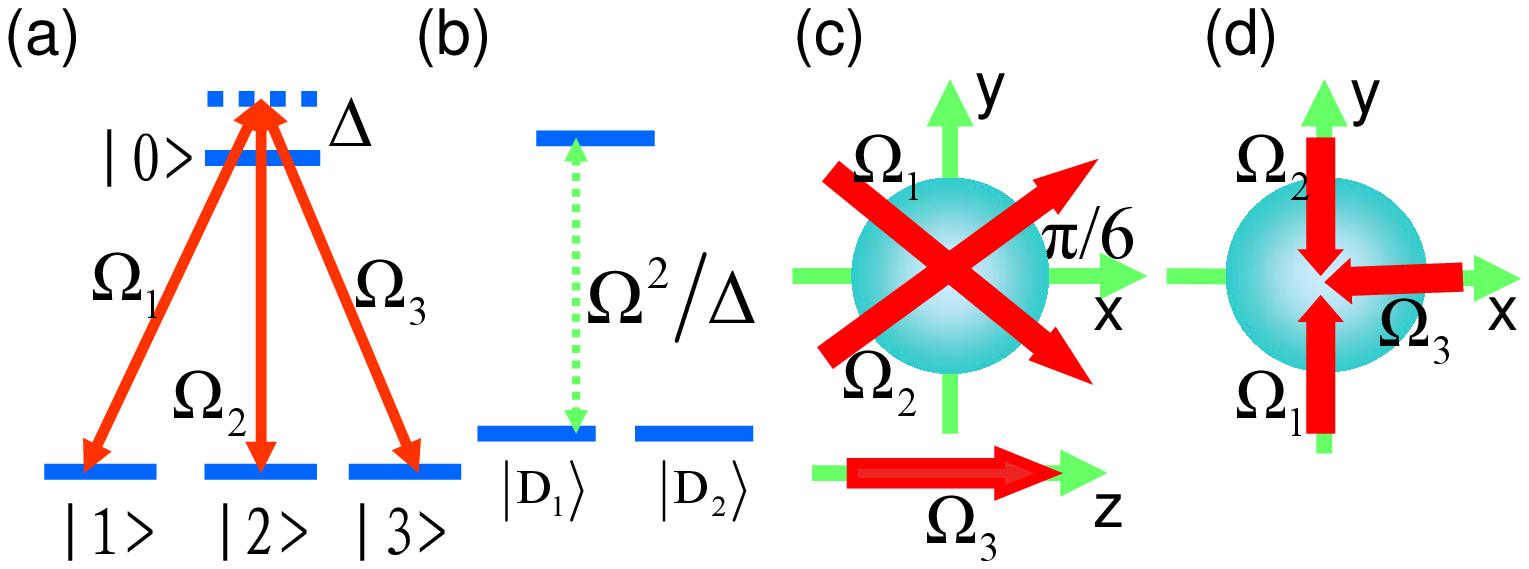} \vspace{-10pt}
\caption{(Color online) A tripod scheme for implementing Rashba spin-orbit
coupling. (a) The atom-laser coupling. (b) The lowest three energy levels
composed of hyperfine ground states $\left\vert 1\right\rangle $, $%
\left\vert 2\right\rangle $, $\left\vert 3\right\rangle $. $\left\vert
D_{1}\right\rangle $ and $\left\vert D_{2}\right\rangle $ are degenerate
dark states. $\Omega =\protect\sqrt{\left\vert \Omega _{1}^{2}\right\vert
+\left\vert \Omega _{2}^{2}\right\vert +\left\vert \Omega
_{3}^{2}\right\vert }$. (c,d) Two different laser configurations
\protect\cite{Juzeliunas,Zhang} for the tripod scheme.}
\label{setup}
\end{figure}

Consider ultra-cold bosonic atoms confined in a quasi-two-dimensional ($xy$
plane) harmonic trap with a tripod electronic level scheme (Fig. \ref{setup}%
a). The atom dynamics along the $\mathbf{\hat{z}}$ direction are frozen by a
deep optical trap or lattice. The hyperfine ground states $\left\vert
1\right\rangle $, $\left\vert 2\right\rangle $, $\left\vert 3\right\rangle $
are coupled with the excited state $\left\vert 0\right\rangle $ using three
blue-detuned lasers with the Rabi frequencies $\Omega _{1}$, $\Omega _{2}$,
and $\Omega _{3}$. The Hamiltonian of the system can be written as $%
H=\sum_{i}H_{s}(\mathbf{r}_{i})+H_{int}$ with the single particle
Hamiltonian
\begin{equation}
H_{s}(\mathbf{r})=\mathbf{p}^{2}/2m+V(\mathbf{r})+H_{I}(\mathbf{r}),
\label{many}
\end{equation}%
where $V(\mathbf{r})=m\left( \omega _{\perp }^{2}r^{2}+\omega
_{z}^{2}z^{2}\right) /2$ is the harmonic trapping potential with the
corresponding trapping frequencies $\omega _{z}$ and $\omega _{\perp }$. $%
H_{I}(\mathbf{r})=-\hbar \Delta \left\vert 0\right\rangle \left\langle
0\right\vert -\hbar \left( \Omega _{1}\left\vert 0\right\rangle \left\langle
1\right\vert +\Omega _{2}\left\vert 0\right\rangle \left\langle 2\right\vert
+\Omega _{3}\left\vert 0\right\rangle \left\langle 3\right\vert +H.c.\right)
$ describes the atom-laser interaction, and $\Delta $ is the detuning to the
state $\left\vert 0\right\rangle $. The diagonalization of the Hamiltonian $%
H_{I}(\mathbf{r})$ yields two degenerate darks states $\left\vert
D_{1}\right\rangle =\sum_{\alpha =1}^{3}d_{\alpha }(\mathbf{r})|\alpha
\rangle $, $\left\vert D_{2}\right\rangle =\sum_{\alpha =1}^{3}f_{\alpha }(%
\mathbf{r})|\alpha \rangle $ and two bright states (Fig. \ref{setup}b),
where the coefficients $d_{\alpha }(\mathbf{r})$, $f_{\alpha }(\mathbf{r})$
are determined by the laser parameters $\Omega _{i}$ and $\Delta $ \cite%
{note1}. In experiments, a large detuning $\Delta $ is chosen to suppress
the spontaneous emission of photons that heats the atom gas. The
blue-detuned lasers are used to ensure the degenerate dark states are the
ground states of the system to avoid the collision loss. In the subspace
spanned by the pseudospin states $\left\vert \uparrow \right\rangle \equiv
\left\vert D_{1}\right\rangle $ and $\left\vert \downarrow \right\rangle
\equiv \left\vert D_{2}\right\rangle $,
\begin{equation}
H_{s}(\mathbf{r})=\mathbf{p}^{2}/2m+\gamma \left( p_{x}\sigma
_{y}-p_{y}\sigma _{x}\right) +V(\mathbf{r}),  \label{single}
\end{equation}%
where $\gamma $, the Rashba SOC strength, is equal to $\hbar k/2\sqrt{3}m$
and $(\sqrt{2}-1)\hbar k/m$ for the two laser setups shown in Figs. \ref%
{setup}c and \ref{setup}d, respectively. $k$ is the wavevector of the
lasers. The \textit{s}-wave scattering interaction between atoms can be
written as $H_{int}=\sum_{i<j}\sum_{\alpha ,\beta =1}^{3}g_{\alpha \beta
}\delta (\mathbf{r}_{i}^{\alpha }-\mathbf{r}_{j}^{\beta })\left\vert \alpha
\right\rangle _{i}\left\vert \beta \right\rangle _{j}\left\langle \alpha
\right\vert _{i}\left\langle \beta \right\vert _{j}$, where $\mathbf{r}%
_{i}^{\alpha }$ is the position of the atom $i$ in the hyperfine state $%
\left\vert \alpha \right\rangle _{i}$, $g_{\alpha \beta }=4\pi \hbar
^{2}a_{\alpha \beta }/m$, $a_{\alpha \beta }$ is the \textit{s}-wave
scattering length between atoms in the hyperfine states $|\alpha \rangle $
and $|\beta \rangle $.

%%%%%%%%%%%%%%%%%%%%%%%%%%%%%%%%%%%%%%%%%%%%%%%%%%%%%%%%%%%%%%%%%%
%%%%%%%%%%%%%%%%%%%%%%%%%%%%%%%%%%%%%%%%%%%%%%%%%%%%%%%%%%%%%%%%%%%

In the Hartree approximation, the many-body wavefunction of the bosonic
system can be taken as the product of the single-particle wave function $%
|\Psi (\mathbf{r}_{1},\mathbf{r}_{2},\ldots ,\mathbf{r}_{N})\rangle
=\prod_{i=1}^{N}|\Phi (\mathbf{r}_{i})\rangle $, where $\left\vert \Phi (%
\mathbf{r})\right\rangle =\Phi _{\uparrow }(\mathbf{r})|D_{1}(\mathbf{r}%
)\rangle +\Phi _{\downarrow }(\mathbf{r})|D_{2}(\mathbf{r})\rangle $ with
the normalization condition $\int d\mathbf{r}(|\Phi _{\uparrow }(\mathbf{r}%
)|^{2}+|\Phi _{\downarrow }(\mathbf{r})|^{2})=1$. The G-P equation can be
obtained through the standard minimization of the mean-field energy
functional $E(\Phi _{\uparrow },\Phi _{\downarrow })=\langle \Psi |H|\Psi
\rangle $ with respect to $\Phi _{\uparrow }^{\ast }$ and $\Phi _{\downarrow
}^{\ast }$ \cite{Pethick}. Because the scattering length is between atoms in
different hyperfine states, the interaction energy $\langle \Psi
|H_{int}|\Psi \rangle $ should be evaluated within the hyperfine state
basis, \textit{i.e.}, $|\Phi (\mathbf{r})\rangle =\sum_{\alpha =1}^{3}[\Phi
_{\uparrow }(\mathbf{r})d_{\alpha }(\mathbf{r})+\Phi _{\downarrow }(\mathbf{r%
})f_{\alpha }(\mathbf{r})]|\alpha \rangle $. With a straightforward
calculation for such energy functional minimization, we derive the G-P
equation for a SOC-BEC
\begin{equation}
i\hbar \partial \Phi /\partial t=H_{s}\Phi +\Gamma \Phi .  \label{nonlinear1}
\end{equation}%
Here the two component wavefunction $\Phi =(\Phi _{\uparrow },\Phi
_{\downarrow })^{T}$ in the pseudospin basis $\left\{ \left\vert \uparrow
\right\rangle ,\left\vert \downarrow \right\rangle \right\} $. The nonlinear
term $\Gamma =%
\begin{pmatrix}
\Gamma _{1} & \Gamma _{2} \\
\Gamma _{2}^{\ast } & \Gamma _{3}%
\end{pmatrix}%
$, where $\Gamma _{1}=C_{1}|\Phi _{\uparrow }|^{2}+C_{2}|\Phi _{\downarrow
}|^{2}+2\text{Re}(C_{3}\Phi _{\uparrow }^{\ast }\Phi _{\downarrow })$, $%
\Gamma _{2}=C_{3}|\Phi _{\uparrow }|^{2}+C_{4}|\Phi _{\downarrow
}|^{2}+C_{5}\Phi _{\uparrow }^{\ast }\Phi _{\downarrow }$, and $\Gamma
_{3}=C_{2}|\Phi _{\uparrow }|^{2}+C_{6}|\Phi _{\downarrow }|^{2}+2\text{Re}%
(C_{4}\Phi _{\uparrow }^{\ast }\Phi _{\downarrow })$ \cite{coe}. $C_{3}$, $%
C_{4}$ and $C_{5}$ terms, which are absent in previous study, originate from
the linear superposition of hyperfine states for a pseudospin state. We see
that not only the density, but also the relative phase between two
components, play an important role on the dynamics of the BEC. Although our
derivation for the mean-field interaction terms is based on the tripod
schemes illustrated in Fig. \ref{setup}, it also applies to other schemes
for generating Rashba SOC using more hyperfine states (e.g. \cite{Ian3}).

For simplicity, we assume $g_{\alpha \beta }=g_{1}$ for $\alpha \neq \beta $
and $g_{\alpha \beta }=g_{0}$ for $\alpha =\beta $ with the corresponding
scattering lengths $a_{1}$ and $a_{0}$. However the nonlinear G-P equation (%
\ref{nonlinear1}) works for any $g_{\alpha \beta }$. $C_{i}$ depend strongly
on the laser configurations for implementing the Rashba SOC. Here we
consider two laser setups for the tripod level scheme (Fig. \ref{setup}a)
that have been investigated previously in the literature \cite%
{Juzeliunas,Zhang,note1}. In the first setup (Fig. \ref{setup}c), we have
the SOC strength $\gamma =$ $k\sqrt{\hbar /12m\omega _{\perp }}$, $%
C_{3}=C_{4}=C_{5}=0$, $C_{1}=C_{6}=\chi (2a_{0}+4a_{1})/3$, and $C_{2}=\chi
(4a_{0}+2a_{1})/3$ (the simple system) with $\chi =2N\sqrt{2\pi m\omega
_{z}/\hbar }$. Here we rescale the G-P equation (\ref{nonlinear1}) with the
energy, time and length units $\hbar \omega _{\perp }$, $\omega _{\perp
}^{-1}$, and $\sqrt{\hbar /m\omega _{\perp }}$ respectively. The unit for $%
\gamma $ is a velocity $\sqrt{\hbar \omega _{\perp }/m}$. In the second
setup (Fig. \ref{setup}d), $\gamma =(\sqrt{2}-1)k\sqrt{\hbar /m\omega
_{\perp }}$, $C_{1}=C_{6}=\chi \lbrack (12-8\sqrt{2})a_{0}+(8\sqrt{2}%
-10)a_{1}]$, $C_{2}=\chi \lbrack (24-16\sqrt{2})a_{0}+(16\sqrt{2}-22)a_{1}]$%
, $C_{3}=C_{4}=-i\chi (7\sqrt{2}-10)(a_{0}-a_{1})$, and $C_{5}=\chi (4\sqrt{2%
}-6)(a_{0}-a_{1})$ (the complex system).

\begin{figure}[t]
\label{wavefunction} \includegraphics[width=1\linewidth]{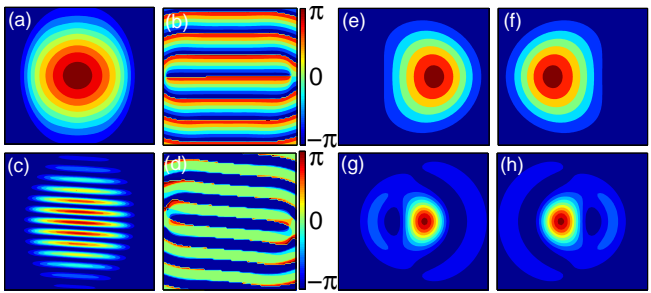} \vspace{%
-10pt}
\caption{(Color online) Condensate wavefunction $\Phi $ in the simple
system. (a,b) Density $\left\vert \Phi _{\uparrow }\right\vert ^{2}$ (a) and
phase arg$\left( \Phi _{\uparrow }\right) $ (b) in the TF phase. $C_{1}=10$,
$C_{2}=6$, $\protect\gamma =10$. (c,d) In the stripe phase. $C_{1}=6$, $%
C_{2}=10$, $\protect\gamma =10$. \ (e,f) Density $\left\vert \Phi _{\uparrow
}\right\vert ^{2}$ and $\left\vert \Phi _{\downarrow }\right\vert ^{2}$ with
$C_{1}=10,C_{2}=6$, $\protect\gamma =1$. (g,h) Density $\left\vert \Phi
_{\uparrow }\right\vert ^{2}$ and $\left\vert \Phi _{\downarrow }\right\vert
^{2}$ with $C_{1}=C_{2}=0,\protect\gamma =10$.}
\label{ground}
\end{figure}

We numerically solve the G-P equation (\ref{nonlinear1}) using the imaginary
time evolution method and obtain the ground state of the condensate. In the
simple system, there are two different types of phases in the region of
strong SOC ($\gamma \gg 1$) and interaction ($C_{i}\gg 1$): the Thomas-Fermi
(TF) phase (Fig. \ref{ground}a) when $C_{1}\geq C_{2}$ (equivalent to $%
a_{1}\geq a_{0}$) and the stripe phase when $C_{1}<C_{2}$ (Fig. \ref{ground}%
c) \cite{Wang}. In the TF phase, the maximum densities of two components
locate at the harmonic trap center, but the phase of the condensate varies
like a plane wave $\exp \left( i\mathbf{k\cdot r}\right) $ (Fig. \ref{ground}%
b). In the stripe phase, the density for each pseudospin component forms a
set of stripes (Fig. \ref{ground}c) and two components are spatially
separated. There is a sharp change of the condensate phase in and outside
the stripe region (Fig. \ref{ground}d). The total density of two components
has a TF distribution for both phases. The direction of the condensate phase
variation $\nabla \arg \left( \Phi _{\sigma }\right) $ is spontaneously
selected, along which the density distribution is wider. We find that the
density profile in each hyperfine state has the same structure (TF or
stripe) as that in the pseudospin state, therefore the TF and stripe phases
should be observable by directly measuring the density in each hyperfine
state.

%%%%%%%%%%%%%%%%%%%%%%%%%%%%%%%%%%%%%%%%%%%%%%%%%%%%%%%%%%%%%%%%%%
%%%%%%%%%%%%%%%%%%%%%%%%%%%%%%%%%%%%%%%%%%%%%%%%%%%%%%%%%%%%%%%%%%%

In the medium SOC region ($\gamma \sim 1$), there exists spatial separation
between the atom densities of two pseudospin components when $C_{1}\geq
C_{2} $ (Figs. \ref{ground}e, \ref{ground}f). The separation can be
understood from the spin-dependent force $\mathbf{F}=\frac{d\mathbf{p}}{dt}%
=2\gamma ^{2}(\mathbf{k}\times \hat{e}_{z})\sigma _{z}$ generated by the
Rashba SOC, where $\mathbf{k}$ is the momentum of atoms and along the
condensate phase variation direction. $\mathbf{F}$ has the opposite
directions for two pseudospins and is along the $\mathbf{\hat{x}}$ direction
when the phase variation is along the $\mathbf{\hat{y}}$ direction (Fig. \ref%
{ground}b), leading to the spatial separation of two pseudospin components
along the $\mathbf{\hat{x}}$ direction (Figs. \ref{ground}e, \ref{ground}f).
The separation due to the spin-dependent force is more transparent in the
non-interacting region (Figs. \ref{ground}g, \ref{ground}h) without
involving the complexity from the interaction. However, the separation
between two components decreases not only for very strong SOC ($\gamma \gg 1$%
, Fig. \ref{ground}a), but also for very weak SOC ($\gamma \ll 1$), which
can be\ understood based on $\mathbf{F}\propto \gamma ^{2}$ and the zero
separation for a regular spinor BEC without SOC.

%%%%%%%%%%%%%%%%%%%%%%%%%%%%%%%%%%%%%%%%%%%%%%%%%%%%%%%%%%%%%%%%%%
\begin{figure}[b]
\includegraphics[width=1\linewidth]{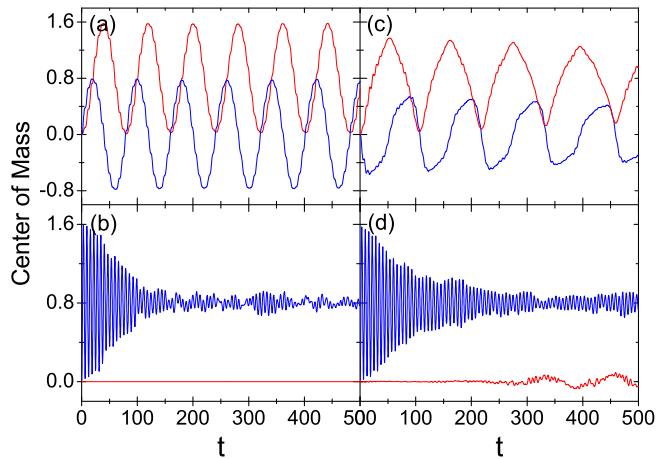} \vspace{-10pt}
\caption{(Color online) The center of mass motion. Red lines: $\langle
x\rangle $; Blue lines: $\langle y\rangle $. (a,b) The simple system with
the corresponding condensate wavefunction in Figs. \protect\ref{ground}a and
\protect\ref{ground}b. The shift of the center of the harmonic trap $D=0.8$.
The shifts are along the $\mathbf{\hat{x}}$ (a) and $\mathbf{\hat{y}}$ (b)
directions respectively. (c,d) The same COM motion as (a,b) but for the
complex system. Parameters are the same as that in (a,b) except that new
terms $C_{3}$, $C_{4}$, $C_{5}$ are included in the condensate and COM
motion caculations.}
\label{COM}
\end{figure}
%%%%%%%%%%%%%%%%%%%%%%%%%%%%%%%%%%%%%%%%%%%%%%%%%%%%%%%%%%%%%%%%%%%

In the complex system, similar TF and stripe phases also exist in the strong
SOC and interaction region. However, significant differences between the
simple and complex systems exist for the low energy collective excitations
even though their condensate phases are similar. Here we consider the COM
motion of the condensate subject to a sudden shift of the center of the
harmonic trap. It is well-known that the COM motion of a BEC without SOC is
a dipole oscillation whose frequency is the harmonic trap frequency and does
not depend on the nonlinearity \cite{Stringari}. On the other hand, the
interference between two Rashba spin-orbit energy bands yields the ZB
oscillation \cite{Vaishnav} for a single particle in the free space with the
oscillation period inversely proportional to $\gamma ^{2}$. In this Letter,
we study the COM motion of a BEC in the presence of interaction, a harmonic
trap, and Rashba SOC.

\begin{figure}[t]
\vspace{5pt} \includegraphics[width=1\linewidth]{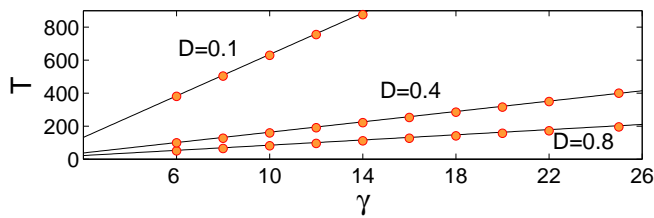} \vspace{-20pt}
\caption{ (Color online) Plot of the COM oscillation period $T$ with respect
to the SOC strength $\protect\gamma $ for the simple system in the TF phase.
Circles are from the numerical stimulation of the G-P equation (\protect\ref%
{nonlinear1}). The lines are from Eq. (\protect\ref{period}).}
\label{Tfig}
\end{figure}

The COM motion%
\begin{equation}
\langle \mathbf{r}(t)\rangle =\int (|\Phi _{\uparrow }(\mathbf{r}%
,t)|^{2}+|\Phi _{\downarrow }(\mathbf{r},t)|^{2})\mathbf{r}d\mathbf{r}
\label{COMM}
\end{equation}%
of the condensate can be obtained by numerically solving the G-P equation (%
\ref{nonlinear1}) and is plotted in Figs. \ref{COM}a and \ref{COM}b (Figs. %
\ref{COM}c, \ref{COM}d) for the TF phase in the simple (complex) system. In
the simple system, when the shift of the center of the harmonic trap $D$ is
along the $\mathbf{\hat{x}}$ direction (perpendicular to the condensate
phase variation $\nabla \arg \left( \Phi _{\uparrow }\right) $ direction $%
\mathbf{\hat{y}}$), the COM motion along the $\mathbf{\hat{x}}$ direction is
perfectly periodic with two periods (Fig. \ref{COM}a): one corresponds to
the harmonic trap frequency, the other is much larger and linearly
proportional to the SOC strength $\gamma $ (Fig. \ref{Tfig}). The COM motion
along the $\mathbf{\hat{y}}$ direction is similar. However, when the shift
of the harmonic trap is along the $\mathbf{\hat{y}}$ direction, the COM
motion along the $\mathbf{\hat{y}}$ direction possesses only the harmonic
trap frequency and the oscillation amplitude is strongly damped (Fig. 3b).
The COM motion along the $\mathbf{\hat{x}}$ direction vanishes.

The different COM motions along different shifting directions may be
understood from the single atom dynamics with the Rashba SOC. The spin-orbit
coupled atoms have two bands with energies $E_{\pm }=\hbar ^{2}k^{2}/2m\pm
\gamma k$ and corresponding wavefunctions $\phi _{\pm }=\exp (i\mathbf{k}%
\cdot \mathbf{r})(1,\pm ie^{i\varphi _{\mathbf{k}}})^{T}/\sqrt{2}$, where $%
\varphi _{\mathbf{k}}=\arg (k_{x}+ik_{y})$. The ground state of the atom $%
\phi _{-}$ stays at the potential minimum located at $k_{0}=\gamma m/\hbar
^{2}$ with the direction of $\mathbf{k}_{0}$ spontaneously selected. The
shift of the harmonic trap corresponds to adding a momentum $\mathbf{p}$
into the ground state, leading to an initial state $\phi _{ini}=\exp (i%
\mathbf{p}\cdot \mathbf{r})\phi _{-}$. The time evolution of the
wavefunction can be written as $\varphi (\mathbf{r},t)=\int d\mathbf{k}%
[A_{-}\exp (iE_{-}t)\phi _{-}+A_{+}\exp (iE_{+}t)\phi _{+}]$, where $A_{\pm
}=\left\langle \phi _{\pm }|\phi _{ini}\right\rangle $. When $\mathbf{p}$%
\textbf{\ }is perpendicular to the direction of $\mathbf{k}_{0}$ (Fig. \ref%
{ground}b), both $A_{\pm }$ are nonzero, and there is an interference
between two spin-orbit coupled bands, leading to the SOC dependent
oscillation of the COM motion (Fig. \ref{COM}a). In contrast, when $\mathbf{p%
}$ is along the direction of $\mathbf{k}_{0}$, one of $A_{\pm }$ must be
zero and there is only one oscillation frequency due to the harmonic trap.

%%%%%%%%%%%%%%%%%%%%%%%%%%%%%%%%%%%%%%%%%%%%%%%%%%%%%%%%%%%%%%%%%%
%%%%%%%%%%%%%%%%%%%%%%%%%%%%%%%%%%%%%%%%%%%%%%%%%%%%%%%%%%%%%%%%%%%

We numerically calculate the new oscillation period $T$ in Fig. \ref{COM}a
for several different sets of interaction parameters, and find that it does
not depend on the interaction strengths $C_{1}$ and $C_{2}$, indicating
essentially single particle physics in this system. In Fig. \ref{Tfig}, we
plot the dependence of $T$ with respect to $\gamma $ for three harmonic trap
shifts $D$. The numerical data can be well fitted with an analytic formula
\begin{equation}
T=2\pi (1+\gamma /D).  \label{period}
\end{equation}%
We see that the period is linearly proportional to $\gamma $ and $D^{-1}$.
In the strong SOC region ($\gamma \gg 1$), the strong SOC couples many
harmonic oscillator states and tends to reduce the single particle energy
splitting, as confirmed by our numerical calculation, resulting in a large
oscillation period. Note that this is very different from the ZB oscillation
in the weak SOC region \cite{Vaishnav}, where the oscillation period is
proportional to $\gamma ^{-2}$ because the energy splitting between two
spin-orbit coupled bands in this region is proportional to $\gamma ^{2}$, as
confirmed in our numerical simulation. Finally, a smaller shift $D$ leads to
less excitations to the high energy states, yielding a larger COM
oscillation period. \ As $\gamma \rightarrow 0$, $T\rightarrow 2\pi $, the
oscillation period for the harmonic trap frequency, as expected. In the weak
and medium interaction or SOC region, our numerical results show that the
oscillation amplitude decays with time due to the phase separation of the
densities of two pseudospin components (Figs. \ref{ground}e-\ref{ground}h)
that leads to the decoherence in the COM motion.

In the stripe phase of the simple system, the COM motion is always damped
because of the decoherence originating from the spatial separation of two
spin components. Similar as the TF phase, the new oscillation period
disappears (emerges) when the shift is along (perpendicular to) the
condensate phase variation direction.

The COM motion is strongly modified in the complex system. In Fig. \ref{COM}%
c and \ref{COM}d, we plot the COM motion in the TF phase. Clearly, the
direction dependence of the new oscillation period is the same as that for
the simple system. However, the oscillation period $T$ is very different
from that in the simple system (Fig. \ref{COM}a) although the complex system
has the same parameters ($C_{1}$, $C_{2}$, $\gamma $) as the simple one. A
damping of the oscillation amplitude is also observed, in contrast to the
perfect oscillation in the simple system. Furthermore, when the shift of the
harmonic trap is along the condensate phase variation direction, the COM
motion perpendicular to the shift direction emerges after a long time (Fig. %
\ref{COM}d), in contrast to the zero motion in the simple system (Fig. \ref%
{COM}b).

In experiments, we can choose the trapping frequency $\omega _{\bot }=2\pi
\times 20Hz$, which yields the time unit $\omega _{\bot }^{-1}=8ms$ and
length unit $a_{h}=\sqrt{\hbar /m\omega _{\bot }}=2.4$ $\mu m$. For Rb
atoms, the spin-orbit coupling strength $\gamma \sim 1-10$. The interaction
strength $C_{1}$, $C_{2}$ can be tuned through the Feshbach resonance or by
adjusting the atom number. With $a_{0},a_{1}\sim 100a_{B}$ for Rb atoms, $%
C_{1}$, $C_{2}\sim 10$ for $N\sim 300$, where $a_{B}=0.53$ \AA\ is the Bohr
radius. We consider a multiple layer system generated by an optical lattice
along the $z$ direction, therefore the total number of atoms is $\sim 10^{4}$%
. We also confirm that the mean field dynamics are similar for larger
interaction strength $C_{1}$, $C_{2}\sim 100$, therefore the total number of
atoms can be $\sim 10^{5}$. For an initial shift of the harmonic trap $D=$ $%
2 $ and $\gamma =6$, we find the center of mass oscillation period $T\sim
200 $ ms, which is much shorter than the lifetime of the BEC and should be
observable in experiments.

In summary, we derive a generic G-P equation and investigate the mean field
dynamics for SOC-BECs. We emphasize that our G-P equation may serve as the
starting point for the future study on the dynamics of SOC-BECs. Our
predicted new oscillation period in the COM motion is observable in
experiments and may provide a powerful tool for exploring the phase of the
condensate wavefunction as well as the low energy excitations in SOC-BECs.

We thank Gang Chen for helpful discussion. This work is supported by the ARO
(W911NF-09-1-0248), DARPA-YFA (N66001-10-1-4025), NSF (PHY-1104546), and
DARPA-MTO (FA955-10-1-0497).

\end{document}